\documentclass{aa}  

\usepackage{graphicx}
\usepackage{txfonts}
\usepackage{xcolor}
\usepackage{hyperref}
\hypersetup{colorlinks, linkcolor=blue, citecolor=blue, urlcolor=blue}
\usepackage{amsmath}
\newcommand{\thetae}{\theta_{\rm E}}

\newcommand{\pie}{\pi_{\rm E}}

\newcommand{\dl}{D_{\rm L}}

\definecolor{brown}{rgb}{0.59, 0.29, 0.0}
\definecolor{darkgreen}{rgb}{0.0, 0.42, 0.24}
\definecolor{darkblue}{rgb}{0.01, 0.31, 0.59}
\definecolor{darkblue}{rgb}{0.0, 0.25, 0.42}
\definecolor{blue}{rgb}{0.0,0.0,1.0}
\definecolor{green}{rgb}{0.0,1.0,0.0}

\begin{document}

\title{KMT-2018-BLG-0748Lb: Sub-Saturn Microlensing Planet Orbiting 
an Ultracool Host }
%\subtitle{I. Overviewing the $\kappa$-mechanism}

\author{
%G. Wuchterl \inst{1} \and 
%C. Ptolemy \inst{2}
%\fnmsep\thanks{Just to show the usage of the elements in the author field} 
% leading author -----------------------------
     Cheongho~Han\inst{1} 
\and In-Gu~Shin\inst{2} 
\and Youn~Kil~Jung\inst{2} 
\and Doeon~Kim\inst{1}
\and Jennifer~C.~Yee\inst{3} 
\\
(Leading authors)\\
and \\
% KMTNet ---------------------------
     Michael~D.~Albrow\inst{4} 
\and Sun-Ju~Chung\inst{2,5}  
\and Andrew~Gould\inst{6,7}
\and Kyu-Ha~Hwang\inst{2} 
\and Chung-Uk~Lee\inst{2} 
\and Yoon-Hyun~Ryu\inst{2} 
\and Yossi~Shvartzvald\inst{8} 
\and Weicheng~Zang\inst{9}
\and Sang-Mok~Cha\inst{2,10} 
\and Dong-Jin~Kim\inst{2} 
\and Hyoun-Woo~Kim\inst{2} 
\and Seung-Lee~Kim\inst{2,5} 
\and Dong-Joo~Lee\inst{2} 
\and Yongseok~Lee\inst{2,10} 
\and Byeong-Gon~Park\inst{2,5} 
\and Richard~W.~Pogge\inst{7}
\\
(The KMTNet Collaboration),\\
}

\institute{
%Institute for Astronomy (IfA), University of Vienna, T\"urkenschanzstrasse 17, A-1180 Vienna\\ \email{wuchterl@amok.ast.univie.ac.at}
%\and
%University of Alexandria, Department of Geography, ...\\ \email{c.ptolemy@hipparch.uheaven.space}
%\thanks{The university of heaven temporarily does not accept e-mails}
     Department of Physics, Chungbuk National University, Cheongju 28644, Republic of Korea                              % (1)
\and Korea Astronomy and Space Science Institute, Daejon 34055, Republic of Korea                                        % (2)
\and Center for Astrophysics $|$ Harvard \& Smithsonian 60 Garden St., Cambridge, MA 02138, USA                          % (3)
\and University of Canterbury, Department of Physics and Astronomy, Private Bag 4800, Christchurch 8020, New Zealand     % (4)
\and Korea University of Science and Technology, 217 Gajeong-ro, Yuseong-gu, Daejeon, 34113, Republic of Korea           % (5)
\and Max Planck Institute for Astronomy, K\"onigstuhl 17, D-69117 Heidelberg, Germany                                    % (6)
\and Department of Astronomy, Ohio State University, 140 W. 18th Ave., Columbus, OH 43210, USA                           % (7)
\and Department of Particle Physics and Astrophysics, Weizmann Institute of Science, Rehovot 76100, Israel               % (8)
\and Department of Astronomy and Tsinghua Centre for Astrophysics, Tsinghua University, Beijing 100084, China            % (9)
\and School of Space Research, Kyung Hee University, Yongin, Kyeonggi 17104, Republic of Korea                           % (10) 
\\
\email{cheongho@astroph.chungbuk.ac.kr}   
}
\date{Received ; accepted}

% \abstract{}{}{}{}{} 
% 5 {} token are mandatory
\abstract
{}
{
We announce the discovery of a microlensing planetary system, in which a sub-Saturn planet 
is orbiting an ultracool dwarf host. 
}
{
We detect the planetary system by analyzing the 
short-timescale ($t_{\rm E}\sim 4.4$~days) lensing event KMT-2018-BLG-0748.  The central part 
of the light curve exhibits asymmetry due to the negative deviations in the rising part and the 
positive deviations in the falling part.
}
{
We find that the deviations are explained by a binary-lens model with a mass ratio between the 
lens components of $q\sim 2\times 10^{-3}$.  The short event timescale together with the small 
angular Einstein radius, $\theta_{\rm E}\sim 0.11$~mas, indicate that the mass of the planet 
host is very small.   The Bayesian analysis conducted under the assumption that the planet 
frequency is independent of the host mass indicates that the mass of the planet is 
$M_{\rm p}=0.18^{+0.29}_{-0.10}~M_{\rm J}$, and the mass of the host, 
$M_{\rm h}= 0.087^{+0.138}_{-0.047}~M_\odot$, is near the star/brown dwarf boundary,  but the 
estimated host mass is sensitive to the assumption about the planet hosting probability.  
High-resolution follow-up observations would lead to revealing the nature of the planet host.
}
{}


\maketitle
%
%-------------------------------------------------------------------

\section{Introduction}\label{sec:one}

As of the time of writing this article, 107 microlensing planets in 100 planetary systems were 
found.\footnote{The Extrasolar Planets Encyclopedia (http://exoplanet.eu)}  Although these 
microlensing planets comprise a minor fraction of all known planets, microlensing provides an 
important method to complement other major planet detection methods because of its capability 
of detecting planets that are difficult to be found by other methods.  See the review 
paper of \citet{Gaudi2012} for various advantages of the microlensing method. Especially, the 
microlensing method enables one to find planets orbiting ultracool dwarfs and brown dwarfs (BDs).  
The most important attribute of the microlensing method to have this capability is that planetary 
microlensing signals rely on the direct gravitational influence of planets and their hosts. As a 
result, it is not needed to measure the host flux for the detection of a planet, and this enables 
the method to detect planets orbiting very faint hosts and even dark objects.

Detecting planets around ultracool dwarfs is important for various reasons. The first importance
lies in the fact that ultracool dwarfs are very common in the Galaxy. According to the present-day
mass function \citep{Chabrier2003}, the number density of stars increases down to the lower stellar
mass limit of $\sim 0.08~M_\odot$. In addition, the extension of the mass function into the BD regime
indicates that BD number density is comparable to the stellar one. Therefore, constructing a
planet sample including those around ultracool dwarfs is essential to fully census the demographics 
of  planets in the Galaxy. Second, planets around ultracool dwarfs provide a test bed to check the 
planet formation scenario. Ultracool dwarfs are the lowest mass objects formed through the process 
of collapsing molecular clouds \citep{Luhman2012}.  The low mass of the central object results in 
the low mass of the accretion disk, and thus the environment of planet formation for ultracool 
dwarfs would be different from that of regular stars. With a sample of planets formed in different 
environment, it would be possible to probe the planet formation process in limiting conditions 
\citep{Payne2007}.

In this paper, we report a microlensing planetary system, in which a sub-Saturn planet orbits an 
ultracool dwarf.   We present the analysis conducted for the planet discovery according to the 
following organization.  In Section~\ref{sec:two}, we describe the observations of the lensing 
event from which the planet is detected.  In Section~\ref{sec:three}, we describe various models 
tested for the interpretation of the data.  In Section~\ref{sec:four}, we explain the procedure of 
determining the source type and the angular radius of the Einstein ring.  In Section~\ref{sec:five}, 
we estimate the mass and distance to the lens.  Discussions on the importance of the discovered 
planetary system and future follow-up observations to refine the physical lens parameters are given 
in Section~\ref{sec:six}.  Summary of the results and conclusion are presented in  
Section~\ref{sec:seven}.

% Figure 1 ------------------------------------------------------
\begin{figure}
\includegraphics[width=\columnwidth]{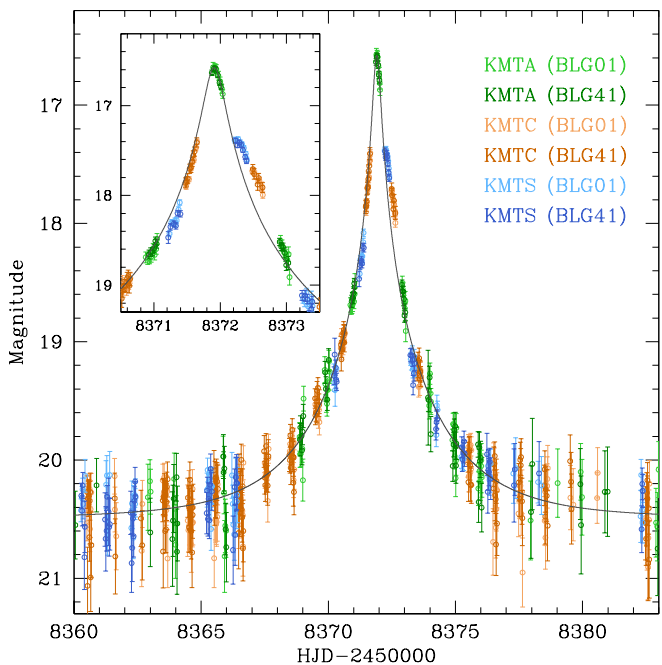}
\caption{
Light curve of the lensing event KMT-2018-BLG-0748. 
The zoomed-in view around the peak is shown in the inset. 
The overplotted curve represents the 1L1S model. 
%\smallskip
}
\label{fig:one}
\end{figure}
% --------------------------------------------------------------

\section{Data}\label{sec:two}

The lensing event from which the planetary system is discovered is KMT-2018-BLG-0748.  The source 
(lensed star) of the event lies toward the bulge field at  $({\rm R.A.}, {\rm decl.})_{\rm J2000} 
=(17 : 51 : 29.88, -30 : 38 : 47.00)$, corresponding to 
$(l, b)=(-0^\circ\hskip-2pt.808,  -1^\circ\hskip-2pt.974)$.  The event reached  its peak  
magnification of $A_{\rm peak} \sim 30$ on 2018-09-10 (${\rm HJD}^\prime \equiv {\rm HJD}-2450000\sim 
8372$).  The event duration, defined by the time span of the source brightening beyond the photometric 
scatter, is about 6~days.

The event was found by the Korea Microlensing Telescope Network \citep[KMTNet:][]{Kim2016} survey. 
The survey commenced its alert-finder system in 2018 \citep{Kim2018b}, but only for a subset of its 
fields that did not include those containing KMT-2018-BLG-0748.  Hence, the event was found from 
the post-season investigation of the data conducted by \citet{Kim2018a}.  The event was in the field 
covered by another lensing survey of the Optical Gravitational Microlensing Experiment 
\citep[OGLE:][]{Udalski2015}, but no trace of lensing signal was detected because of the source 
location at the edge of a camera chip.

Although the event lasted a short period of time, the light curve was continuously and densely 
covered.  This coverage was possible thanks to two main reasons.  First, the event was observed 
using multiple telescopes that were globally distributed in three continents.  The individual 
KMTNet telescopes are located at the Siding Springs Observatory (KMTA) in Australia, the Cerro 
Tololo Inter-American Observatory (KMTC) in South America, and the South African Astronomical 
Observatory (KMTS) in Africa.  The three telescopes are identical with a 1.6~m aperture.  The 
$9{\rm k}\times 9{\rm k}$ mosaic camera mounted on each telescope yields a very wide field of 
view of 4~deg$^2$, and this enables high-cadence observations of the event.  Second, the source 
is positioned in the overlapping region of the two fields, for which the observational cadence 
was highest among the total 27 KMTNet fields.   The two fields, BLG01 and BLG41, were laid out 
to overlap each other in most covered area, with a small offset between the fields to cover the 
gaps among the camera chips.  The individual fields were monitored with a 30~min cadence, resulting 
in a combined cadence of 15~min.  Observations were primarily carried out with the $I$ band, and 
a subset of $V$-band images were acquired to measure the color of the source star.

Photometry of data are conducted using the KMTNet pipeline \citep{Albrow2009}, that utilizes 
the difference imaging method developed by \citet{Tomaney1996} and \citet{Alard1998}.  For a 
subset of the data, we conduct an extra photometry with the pyDIA code \citep{Albrow2017} to 
estimate the color of the source star.  In Section~\ref{sec:four}, we describe the detailed 
procedure of estimating the source color.  We rescale the error bars of the data estimated from 
the photometry pipeline using the method of \citet{Yee2012}.

\section{Interpretation of light curve}\label{sec:three}

\subsection{Single-lens (1L1S) model}\label{sec:three-one}

The lensing light curve of KMT-2018-BLG-0748 is shown in Figure~\ref{fig:one}.  At a casual 
glance, the light curve seems to be that of a standard event involved with a single lens and 
a single source (``1L1S''), and thus we first model the light curve under the 1L1S interpretation.  
Three parameters characterize a 1L1S lensing light curve.  These parameters are $t_0$, $u_0$, and 
$t_{\rm E}$, which represent the time of the minimum lens-source (projected) separation, the 
separation between the lens and source (scaled to the angular Einstein radius $\thetae$) at $t_0$ 
(impact parameter), and the event timescale, respectively.

The 1L1S model curve is plotted over the data points in Figure~\ref{fig:one}.  The modeling is 
conducted so that the peak of the model matches the observed peak at $t_0\sim 8371.9$.  From the 
comparison of the observed data with the model, it appears that the observed light curve is 
approximately described by a 1L1S model with an event timescale of $t_{\rm E}\sim 4$~days.  
However, a close inspection reveals that there exist slight deviations from the model, especially 
in the region around the peak.  See the inset showing the enlarged view of the light curve around 
the peak.

% Figure 2 ------------------------------------------------------
\begin{figure*}
\centering
\includegraphics[width=14.2cm]{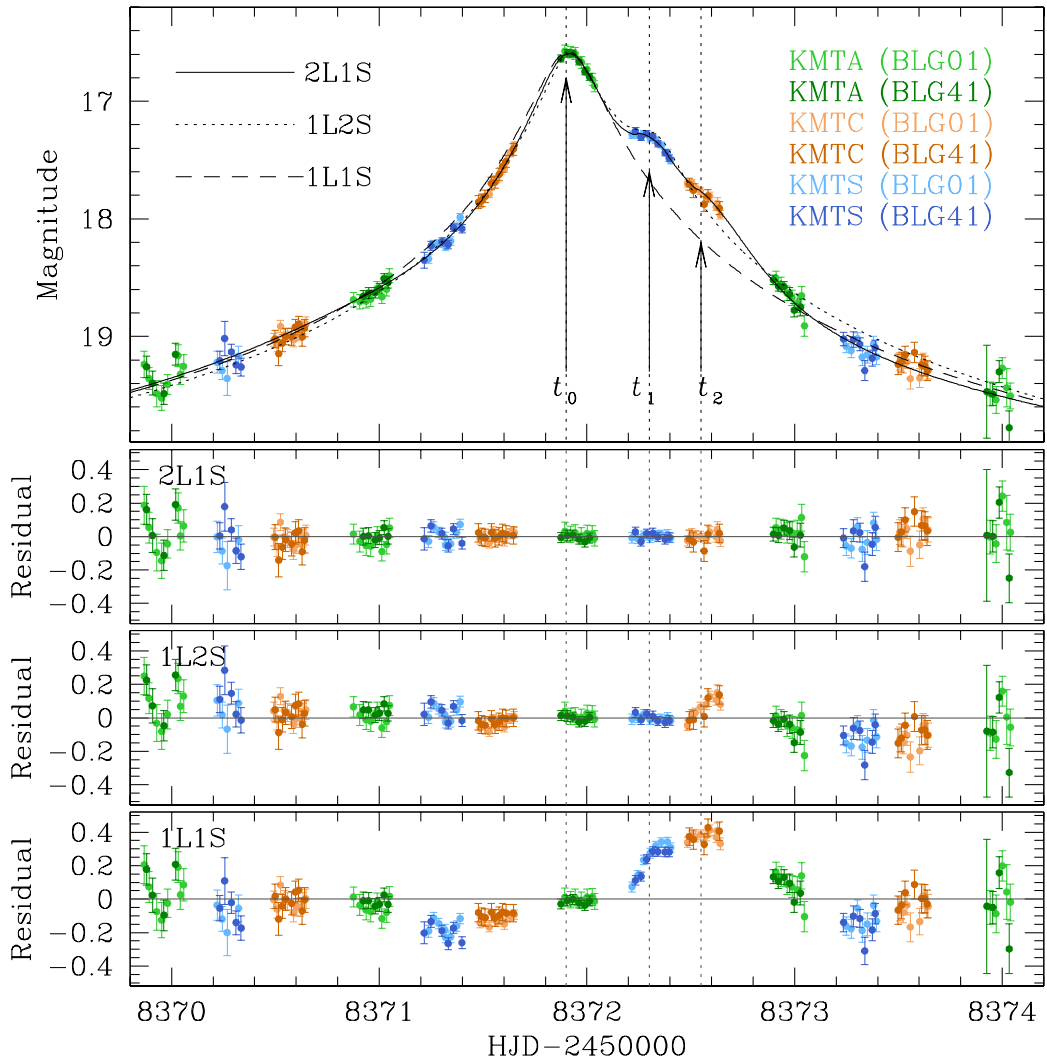}
\caption{
Various tested models and their residuals. 
The top panel shows the three tested models, 2L1S (solid curve), 1L2S (dotted curve), and 1L1S 
(dashed curve) models, and the three lower panels show the residuals from the individual models.  
The three times marked at $t_0\sim 8371.9$, $t_1\sim 8372.3$, and $t_2\sim 8372.6$ represent the 
peak time, and the two bumps, respectively.
%\smallskip
}
\label{fig:two}
\end{figure*}
% --------------------------------------------------------------

In order to inspect the detailed structure of the deviations, we display the peak region of the 
light curve in Figure~\ref{fig:two}.  We also show the residuals from the 1L1S model in the bottom 
panel.  From the inspection of the residuals, it is found that the deviations occur in two major 
parts. The first part is the region in the rising side of the light curve before the peak during 
$8371.2 \lesssim {\rm HJD}^\prime \lesssim 8371.7$. The data points in this region exhibit negative 
deviations with respect to the 1L1S model. The second part is the region in the falling side after 
the peak during $8372.2 \lesssim {\rm HJD}^\prime \lesssim 8373.0$. In contrast to the first part, 
the data points in this region exhibit positive deviations.  To be noted is that the deviations in 
this region exhibit two bumps at $t_1\sim 8372.3$ and $t_2\sim 8372.6$, and the positive deviations 
are followed by slight negative deviations during $8373.2 \lesssim {\rm HJD}^\prime \lesssim 8373.5$. 
The combination of the negative deviations in the rising side and the positive deviations in the 
falling side makes the light curve appear to be asymmetric.

A lensing light curve may become asymmetric due to two major reasons. The first is the binarity of the 
lens and the other is the binarity of the source.  Acceleration of the observer induced by the 
orbital motion of the Earth can also cause a light curve to appear asymmetric, but for short-timescale 
events like KMT-2018-BLG-0748, these microlens-parallax effects \citep{Gould1994} cannot be the cause 
of the light curve asymmetry because the deviation of the source motion from rectilinear during the 
short duration of the lensing magnification is negligible.

% Table 1 ------------------------------------------------
%\begin{deluxetable}{lc}
%\begin{table}{lc}
\begin{table}[htb]
\centering
%\tablecaption{Lesing parameters of 1L2S model\label{table:one}}
\caption{Lesing parameters of 1L2S model\label{table:one}}
%\tablewidth{240pt}
%\tabletypesize{\small}
%\tablehead{
%\multicolumn{1}{c}{Parameter}     &
%\multicolumn{1}{c}{Value}           
%}
%\begin{tabular}{lc}
\begin{tabular*}{\columnwidth}{@{\extracolsep{\fill}}lc}
\hline\hline
\multicolumn{1}{c}{Parameter}     &
\multicolumn{1}{c}{Value}         \\
%\startdata                                              
\hline
$\chi^2$                         &  $1145.8             $     \\
$t_{0,1}$ ({\rm HJD}$^\prime$)   &  $8371.918 \pm 0.004 $     \\
$t_{0,2}$ ({\rm HJD}$^\prime$)   &  $8372.336 \pm 0.006 $     \\
$u_{0,1}$                        &  $0.052 \pm 0.002    $     \\
$u_{0,2}$                        &  $0.033 \pm 0.002    $     \\
$t_{\rm E}$ (day)                &  $3.54 \pm 0.11      $     \\
$q_F$                            &  $0.20 \pm  0.02     $     \\
\hline
%\enddata                            
\end{tabular*}
%\tablecomments{
\tablefoot{
${\rm HJD}^\prime = {\rm HJD}- 2450000$. 
%\smallskip
}
%\end{deluxetable}
\end{table}
% --------------------------------------------------------

\subsection{Binary-source (1L2S) model}\label{sec:three-two}

To investigate the cause of the deviations from the 1L1S model, we first test a model in which 
the source is composed of two stars (``1L2S'' model). Adding one more source component to the lens 
modeling requires one to include additional parameters together with the 1L1S parameters, i.e.,  
$(t_0, u_0, t_{\rm E})$. Following the parameterization of \citet{Hwang2013}, these parameters are 
$t_{0,2}$, $u_{0,2}$, and $q_F$, representing the peak time and impact parameters of the second 
source, and the flux ratio of the second source to the first source, respectively.  To designate 
the peak time and impact parameter associated with the primary source, we use the notations $t_{0,1}$ 
and $u_{0,1}$, respectively.  Modeling is done using the initial parameters estimated from the 1L1S 
modeling and considering the deviation features.

\begin{figure}
\includegraphics[width=\columnwidth]{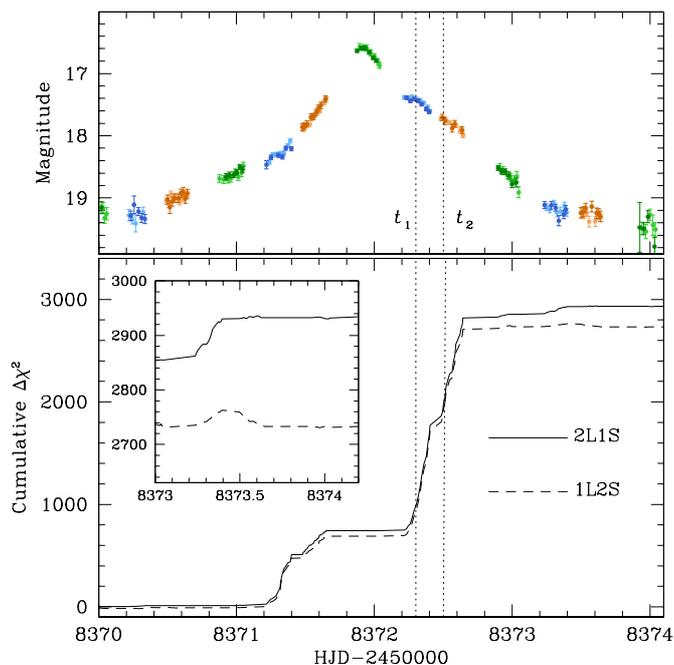}
\caption{
Cumulative $\Delta\chi^2$ distributions of the 2L1S (solid curve) and 1L2S (dashed curve) models 
with respect to the 1L1S model.  The observed light curve is shown in the upper panel to see the 
locations around which the fit improves.  The zoomed-in view of the 2L1S and 1L2S distributions in 
the region of $8373.0 \lesssim {\rm HJD}^\prime \lesssim 8374.2$ is shown in the inset of the lower 
panel.  The times marked by $t_1$ and $t_2$ corresponds to the times of the two bumps marked in 
Fig.~\ref{fig:two}.
%\smallskip
}
\label{fig:three}
\end{figure}

The lensing parameters of the 1L2S model, together with the $\chi^2$ value of 
the fit, are listed in Table~\ref{table:one}.  The model curve (dotted curve in the top panel) 
and residuals (presented in the third panel labeled as ``1L2S'') from the model are shown in 
Figure~\ref{fig:two}.  Comparison of the model fits indicates that the 1L2S model provides a 
substantially better fit than the 1L1S model by $\Delta\chi^2=2739.6$.  To inspect how the 
introduction of a source companion improves the fit, we construct the cumulative distribution of 
$\Delta\chi^2$ between the 1L2S and 1L1S models and present the distribution in Figure~\ref{fig:three}.  
According to the 1L2S model, the second source, which is fainter than the first source by 
$\Delta I=-2.5 \log q_F\sim 1.75$ magnitude, trails the primary source with a time gap of 
$\Delta t=t_{0,2}-t_{0,1}\sim 0.42$~days ($\sim 10$~hours) and approaches the lens with a smaller 
impact parameter ($u_{0,2}\sim 0.033$) than that of the primary source ($u_{0,1}\sim 0.052$).  We 
consider finite-source effects to investigate whether the lens passes over the surfaces of the source 
stars, but it is found that the impact parameters of both source trajectories are greater than the 
source radii, and thus there is no trace of finite-source effects.  The introduction of a binary 
source substantially reduces the negative residuals in the rising side of the light curve.  The 
1L2S model also reduces the positive residuals in the the falling side.  However, the model leaves 
noticeable deviations around the bump at $t_2$ and in the negative deviation region during  
$8373.0 \lesssim {\rm HJD}^\prime \lesssim 8373.5$.

% Figure 3 ------------------------------------------------------
% --------------------------------------------------------------

% Table 2 ------------------------------------------------
\begin{table}[thb]
\centering
\caption{Lesing parameters of 2L1S model\label{table:two}}
\begin{tabular*}{\columnwidth}{@{\extracolsep{\fill}}lc}
\hline\hline
\multicolumn{1}{c}{Parameter}     &
\multicolumn{1}{c}{Value}         \\  
\hline
$\chi^2$                      &  $937.6              $     \\
$t_0$ ({\rm HJD}$^\prime$)    &  $8371.901 \pm 0.004 $     \\
$u_0$                         &  $0.034 \pm 0.002    $     \\
$t_{\rm E}$ (day)             &  $4.38 \pm 0.15      $     \\
$s$                           &  $0.939 \pm 0.003    $     \\
$q$ ($10^{-3}$)               &  $2.03 \pm 0.15      $     \\
$\alpha$ (rad)                &  $-0.023 \pm 0.008   $     \\
$\rho$ ($10^{-3}$)            &  $10.89 \pm 0.97     $     \\
\hline
\end{tabular*}
\tablefoot{
${\rm HJD}^\prime = {\rm HJD}- 2450000$. 
}
\end{table}
% --------------------------------------------------------

The validity of the 1L2S model can be additionally checked with the use of the $V$-band data, 
because the $V$-band data points near the maximum magnification of the second source would
manifest a color change \citep{Gaudi1998, Hwang2013}.  Unfortunately, this test cannot be done 
because the peak region of the second source magnification is sparsely covered by the $V$-band 
data, and the photometry quality of the few data points is not good enough to securely measure 
the color change induced by the binary source.

\subsection{Binary-lens (2L1S) model}\label{sec:three-three}

We also test a model in which the lens is a binary (``2L1S'' model). Similar to the 1L2S modeling,
considering one more lens component requires one to add extra parameters to the lens modeling.  These 
parameters are the $s$, $q$, and $\alpha$, and they denote the projected separation between the lens 
components $M_1$ and $M_2$, the mass ratio, i.e., $q=M_2/M_1$, and the source trajectory angle measured 
from the $M_1$--$M_2$ axis, respectively.  A binary lens induces caustics, at which a point-source 
lensing magnification becomes infinite, and the lensing light curve of a caustic-crossing event deviates 
from that of a point-source event due to the effect of a finite source size.  We consider this finite-source 
effect in modeling by adding one more parameter of $\rho$, which is defined as $\rho=\theta_*/\thetae$  
(normalized source radius).  Here $\theta_*$ denotes the angular size of the source radius.  Finite-source 
magnifications are computed using the numerical method based on the ray-shooting algorithm \citep{Dong2006}.   
In this process, we take account of the surface brightness variation of the source caused by limb darkening.  
We assume that the surface brightness linearly decreases and choose the limb-darkening coefficients 
considering the spectral type of the source.  We discuss the detailed procedure of the source type 
determination in Section~\ref{sec:four}.  With the surface-brightness model $S\propto 1-\Gamma(1-1.5\cos\psi)$, 
we adopt  $(\Gamma_V, \Gamma_I)=(0.62, 0.49)$ for the coefficients in $V$ and $I$ bands, respectively.  
Here $\psi$ represents the angle between two vectors, with origin at the source center and pointing 
toward the observer and to the source surface.

In the 2L1S modeling, the lensing parameters are categorized into two groups.  The parameters $(s, q)$ 
in the first group are searched for using a grid approach because  a lensing magnification can change 
dramatically with a minor variation of the parameters.  For the other parameters, i.e., 
$(t_0, u_0, t_{\rm E}, \alpha, \rho)$, in contrast, the magnification variation is smooth, and thus we 
look for these parameters using a downhill method based on the Markov Chain Monte Carlo (MCMC) algorithm.  
The grid search for the parameters $(s, q)$ is useful because investigating the $\Delta\chi^2$ map on the 
plane of these parameters enables one to find possible degenerate solutions, which yield similar models 
despite the large differences in the lensing parameters.  For KMT-2018-BLG-0748, we identify a unique 
solution without any degeneracy.  With the rough position of the local in the $s$--$q$ plane, we then 
conduct another modeling to find a refined solution by setting, in this time, all lensing parameter, 
including $s$ and $q$, as  free parameters.

The model curve (solid curve in the top panel) obtained from the the 2L1S modeling and the residuals 
(second panel labeled as ``2L1S'') from the model are shown in Figure~\ref{fig:two}.  The lensing 
parameters of the model are presented in Table~\ref{table:two} together with the $\chi^2$ value of 
the fit.  To be noted among the parameters is that the estimated value of $q\sim 2\times 10^{-3}$ is 
very low, suggesting that the companion to the lens is very likely to be a planet.

% Figure 4 ------------------------------------------------------
\begin{figure}
\includegraphics[width=\columnwidth]{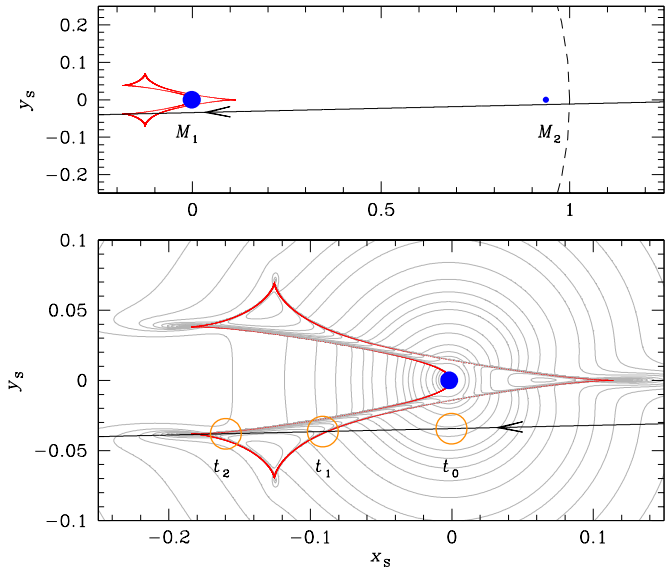}
\caption{
Configuration of the lens system for the 2L1S solution.  The upper panel shows the zoomed-out region 
encompassing the binary lens components, blue dots marked by $M_1$ and $M_2$, and the lower panel 
shows the zoomed-in view of the central magnification region.  The line with an arrows indicates the 
source trajectory, the cuspy closed curve is the caustic, and the dashed circle in the upper panel 
represents the Einstein ring.  In the lower panel, the three empty circles on the trajectory of the 
source indicate the source locations corresponding to the times of $t_0$, $t_1$, and $t_2$ marked 
in Fig.~\ref{fig:two}.  The circle size is scaled to $\thetae$.  The grey curves in the lower panel 
represent equi-magnification contours.  
\smallskip
}
\label{fig:four}
\end{figure}
% --------------------------------------------------------------

We find that the 2L1S model well explains the observed light curve.  The relative goodness of the 2L1S 
model over the other models is shown in the cumulative $\Delta\chi^2$ distribution presented in 
Figure~\ref{fig:three}.  The 2L1S model improves the fit by $\Delta\chi^2=2947.0$ with respect to the 
1L1S model.  The model fit is better than that of the 1L2S model by $\Delta\chi^2=208.2$, and the 
residuals from the 1L2S models, i.e., the bump at around $t_2$ and the negative deviations during 
$8373.0 \lesssim {\rm HJD}^\prime \lesssim 8373.5$, vanish. This indicates that the observed deviations 
are explained by a lens companion rather than a source companion.

% Figure 5 ------------------------------------------------------
\begin{figure}
\includegraphics[width=\columnwidth]{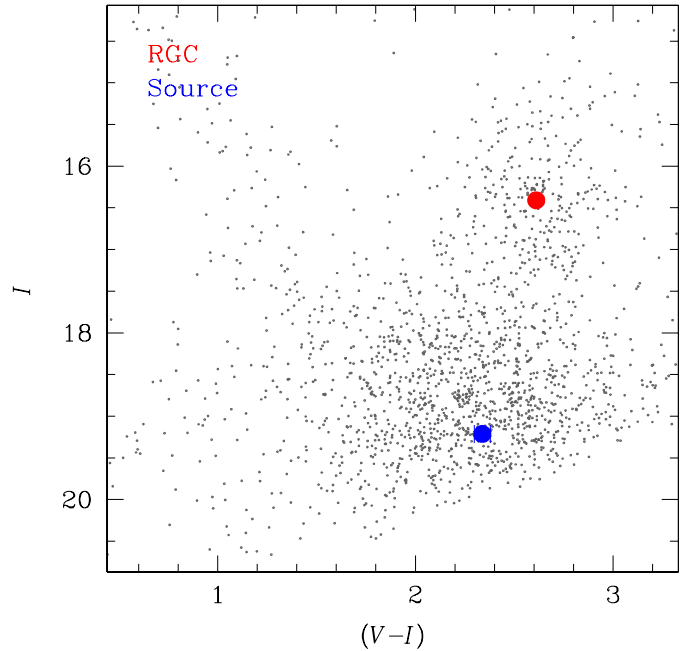}
\caption{
Color-magnitude diagram of stars located within $2^\prime \times 2^\prime$ region around the source.
The blue and red dots represent the positions of the source and RGC centroid, respectively. 
%\smallskip
}
\label{fig:five}
\end{figure}
% --------------------------------------------------------------

The configuration of the lens system according to the 2L1S solution is provided in Figure~\ref{fig:four}.  
In the figure, locations of the binary lens components are marked by blue dots, the source trajectory is 
represented by a line with an arrow, and the caustic is shown in red color (cuspy closed curve).  We note 
that the separation between $M_1$ and $M_2$, $s\sim 0.94$, is similar to $\thetae$, and this results in 
a single resonant caustic.  The caustic has two prongs extending from the position of the primary lens.  
The source moves along the path that is nearly parallel to the $M_1$--$M_2$ axis, approaching the primary 
lens, and crossing the lower prong of the caustic two times, i.e., caustic entry and exit.  The time of the 
source approach to the primary lens corresponds to the time of the peak magnification, $t_0$, and the times 
of the caustic crossings correspond to the times of the bumps in the light curve at $t_1$ and $t_2$ marked 
by arrows in Figure~\ref{fig:two}.  A caustic crossing, in general, results in a rapid rising of the lensing 
magnification, yielding a spike feature in lensing light curves.  For KMT-2018-BLG-0748, however, finite-source 
effects are severe, and thus the caustic crossings result in minor bumps instead of spike features.  According 
to the 2L1S model, the slight negative deviations from the 1L1S model during $8372.2 \lesssim {\rm HJD}^\prime 
\lesssim 8373.0$ are explained by the excess magnification extending from the tip of the caustic.  We note 
that these negative deviations and the bump at $t_2$ cannot be explained by the 1L2S model.

% Table 3 ------------------------------------------------
\begin{table}[thb]
\centering
\caption{Estimated values of $\theta_*$, $\thetae$ and $\mu$ \label{table:three}}
\begin{tabular*}{\columnwidth}{@{\extracolsep{\fill}}lc}
\hline\hline
\multicolumn{1}{c}{Parameter}     &
\multicolumn{1}{c}{Value}         \\  
\hline
$\theta_*$ ($\mu$as)     &  $1.21 \pm 0.10   $     \\
$\thetae$ (mas)          &  $0.111 \pm 0.010 $     \\
$\mu$ (mas~yr$^{-1}$)    &  $9.24 \pm 0.80   $     \\
\hline
\end{tabular*}
%\tablefoot{ ${\rm HJD}^\prime = {\rm HJD}- 2450000$.  }
\end{table}
% --------------------------------------------------------

\section{Angular Einstein radius}\label{sec:four}

We estimate $\thetae$ from the measured value of $\rho$ by $\thetae=\theta_*/\rho$.  For this, we 
first estimate $\theta_*$ based on the de-reddended color, $(V-I)_0$, and brightness, $I_0$, of the 
source using the method of \citet{Yoo2004}.  According to the procedure of this method, we locate the 
source in the instrumental color-magnitude diagram (CMD), measure the offsets of the source in color, 
$\Delta (V-I)$, and brightness, $\Delta I$, from the centroid of the red giant clump (RGC), and then 
estimate $(V-I)_0$ and $I_0$ using the relation
\begin{equation}
(V-I, I)_0 = (V-I, I)_{{\rm RGC,0}} + \Delta (V-I, I). 
\label{eq1}
\end{equation}
Here $(V-I, I)_{{\rm RGC},0}=(1.06, 14.47)$ denote the known values of the RGC centroid from 
\citet{Bensby2013} and \citet{Nataf2013}, respectively.

Figure~\ref{fig:five} shows the source and RGC centroid in the CMD of stars in the vicinity of the 
source.  We note that the CMD is produced by conducting photometry on the KMTC data set using the 
pyDIA software,  but the color and brightness are scaled to those of the OGLE-III system 
\citep{Szymanski2011}  to present the calibrated color and brightness.  The measured positions of 
the source and RGC centroid are 
$(V-I, I)=(2.34\pm 0.04, 19.21\pm 0.03)$ and $(V-I, I)_{\rm RGC}= (2.61, 16.41)$, 
respectively.  By measuring the offsets of  
$\Delta (V-I, I)=(-0.27, 2.80)$, we estimate that the source has a de-reddened color and a brightness of 
\begin{equation} 
(V-I, I)_0 = (0.79 \pm 0.04, 17.27\pm 0.03).  
\label{eq2} 
\end{equation} 
From the measured color and brightness, it is found that the source is a turn-off star with a spectral 
type of late G.

The angular radius of the source is deduced from its color and brightness.  For this, we first 
estimate $V-K$ from the measured $V-I$ using the relation between the two colors \citep{Bessell1988}, 
and then estimate $\theta_*$ using the relation between $(V-K)$ and $\theta_*$ \citep{Kervella2004}.  
This process yields the source radius of
\begin{equation}
\theta_* = 1.21 \pm 0.10~\mu{\rm as}. 
\label{eq3}
\end{equation}
With the estimated $\theta_*$, the Einstein radius and the relative lens-source proper motion are 
estimated by
\begin{equation}
\thetae = {\theta_*\over \rho} = 0.111 \pm 0.013~{\rm mas},
\label{eq4}
\end{equation}
and
\begin{equation}
\mu = {\thetae\over t_{\rm E}} = 9.24 \pm  1.11~{\rm mas}~{\rm yr}^{-1}, 
\label{eq5}
\end{equation}
respectively.  The values of $\theta_*$, $\thetae$, and $\mu$ are summarized in Table~\ref{table:three}.  
Considering that typical lensing events,  generated by M dwarfs located roughly halfway between the 
observer and source, results in $\thetae\sim 0.5$~mas, the measured $\thetae$ value is significantly 
smaller than those of typical events.  Because $\thetae\propto M^{1/2}$, the small value of $\thetae$ 
suggests that the lens is a low-mass object.  In contrast, the measured relative proper motion,
$\sim 9.2~{\rm mas}~{\rm yr}^{-1}$, is considerably higher than a typical value of 
$\sim 5~{\rm mas}~{\rm yr}^{-1}$.

\section{Physical lens parameters}\label{sec:five}

The physical lens parameter of the mass, $M$, and distance, $D_{\rm L}$, are unambiguously determined 
by measuring the two observables of $\thetae$ and $\pie$, i.e.,
\begin{equation}
M={\thetae\over \kappa \pie};\qquad
\label{eq6}
\end{equation}
and
\begin{equation}
D_{\rm L} = {{\rm au}\over \pie\thetae+ \pi_{\rm S}}.
\label{eq7}
\end{equation}
Here $\pie$ is the microlens parallax,  $\kappa =4G/(c^2{\rm au})$, and $\pi_{\rm S}$ denotes the parallax 
to the source \citep{Gould2000b}.  Among the two observables needed to determine $M$ and $D_{\rm L}$, 
$\thetae$ is measured, but $\pie$ is not measured in the case of KMT-2018-BLG-0748, and this makes it difficult 
to determine the physical parameters using the relations in Equations~(\ref{eq6}) and (\ref{eq7}).  We, therefore, 
estimate $M$ and $\dl$ from a Bayesian analysis using the priors for the Galactic model and lens mass function 
together with the measured values of $t_{\rm E}$ and $\thetae$.

% Figure 6 ------------------------------------------------------
\begin{figure}
\includegraphics[width=\columnwidth]{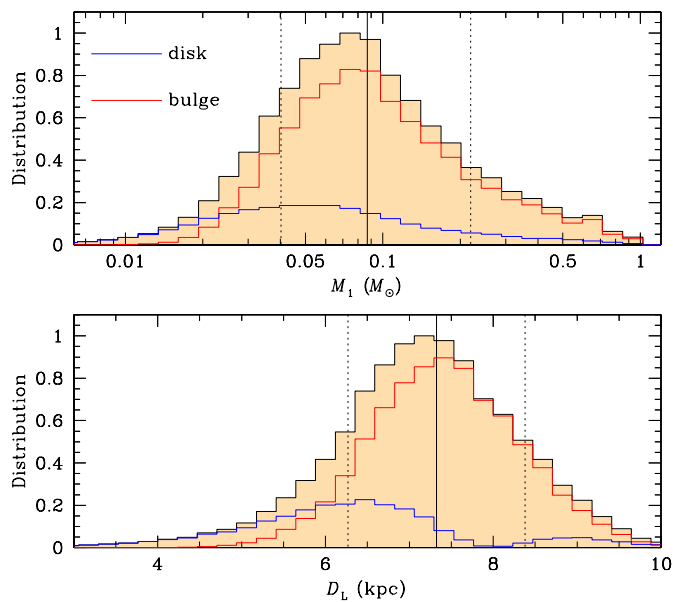}
\caption{
Bayesian posteriors for the host mass (upper panel) and distance (lower panel) estimated from the 
Bayesian analysis. In each panel, the blue and red curves are the distributions contributed by disk 
and bulge lenses, respectively, and the black curve is that of the two lens populations combined.  
The solid and dotted vertical lines indicate the median and $1\sigma$ range, respectively.
\smallskip
}
\label{fig:six}
\end{figure}
% --------------------------------------------------------------

% Figure 7 ------------------------------------------------------
\begin{figure*}
\centering
\includegraphics[width=13cm]{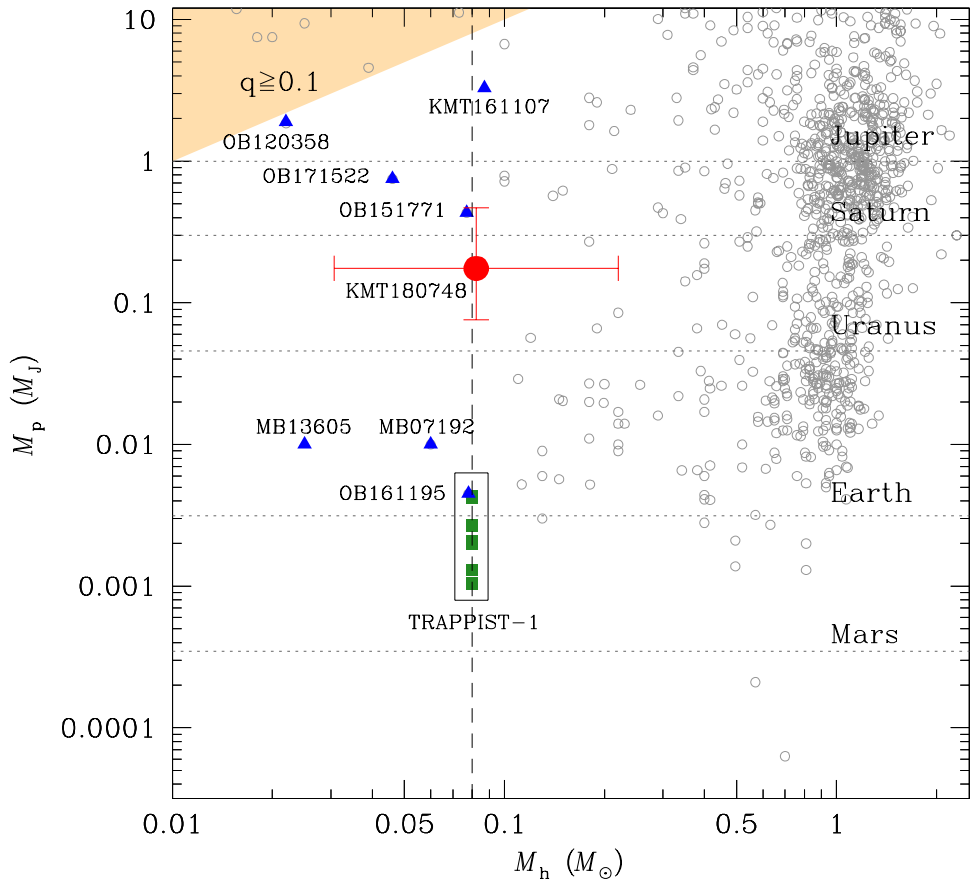}
\caption{
Distribution 
of planets in the plane of the host, $M_{\rm h}$, and planet, $M_{\rm p}$, masses for planetary systems 
with known masses.  Planetary systems with host masses at around or below the stellar mass limit are 
marked by colored points: a red dot for KMT-2018-BLG-0748L, blue triangle dots for the other microlensing 
planetary systems, and green square dots for the seven planets in TRAPPIST-1 system.  The yellow shaded 
area represents the region with mass ratios $q=M_{\rm p}/M_{\rm h}\geq 0.1$. The dashed vertical line represents 
the star/BD boundary. The dotted horizontal lines represent the masses of the Jupiter, Saturn, Uranus, 
Earth, and Mars of the Solar system from the top to bottom.
\smallskip
}
\label{fig:seven}
\end{figure*}
% --------------------------------------------------------------

The Bayesian analysis is conducted by generating numerous ($4\times 10^7$) artificial lensing events from 
a Monte Carlo simulation based on the prior models.  In the simulation, lenses are physically distributed 
following a modified \citet{Han2003} model and the lens-source transverse speeds of events are assigned 
based on the modified dynamical distributing model of \citet{Han1995}.  Compared to the original 
\citet{Han2003} model, in which the Galaxy disk is modeled by a simple double-exponential disk, the disk 
in the modified model has a form of \begin{equation}
\rho_{\rm disk}=\rho_{\rm disk,0} \left[
e^{- (b^2+a^2/h_{R+}^2)^{1/2}} - 
e^{- (b^2+a^2/h_{R-}^2)^{1/2}} 
\right],
\label{eq8}
\end{equation}
by adopting the disk model of \citet{Bennett2014}.  Here $\rho_{\rm disk,0}=1.1~M_\odot/{\rm pc}^3$ 
represents the matter density in the solar neighborhood, $a=[R^2+(z/h_z)^2]^{1/2}$, $h_z=79~{\rm pc}$, 
$b=0.5$, $(h_{R+}, h_{R-})=(2530, 1320)~{\rm pc}$, and $(R,z)$ represent the position in the Galactocentric 
cylindrical coordinates.  This modification is done to prevent the increase of the disk density all the way 
to the Galactic center.  The dynamical model of disk objects is modified to account for the change in the matter
distribution by changing the velocity dispersions along the $(R,z)$ directions as $\sigma_R=\sigma_{R,0} 
[\Sigma(D_{\rm L})/\Sigma(D_{\rm L}=0)]^{1/2}$ and $\sigma_z=\sigma_{z,0} [\Sigma(D_{\rm L})/
\Sigma(D_{\rm L}=0)]^{1/2}$, where $(\sigma_{R,0},\sigma_{z,0})=(30, 20)~{\rm km}~{\rm s}^{-1}$ 
represent the velocity dispersions in the solar neighborhood along the $(R,z)$ directions, respectively.  
For the bulge velocity dispersions, we use the mean values measured by {\it Gaia}.  Considering the short 
timescale and small $\thetae$ of KMT-2018-BLG-0748, lens masses are assigned using the mass function model 
of \citet{Zhang2019}, which extends down to a substellar mass regime.  In the mass function model, we consider 
remnant lenses, i.e., white dwarfs, neutron stars, and black holes, by adopting the \cite{Gould2000a} model.   
We compute the posteriors for $M$ and $D_{\rm L}$ by obtaining the probability distributions of events with 
values of $t_{\rm E}$ and $\thetae$ located within the uncertainty ranges of the measured $t_{\rm E}$ and 
$\thetae$.  We, then, use the median as a representative value and the 16\% and 84\% ranges of the 
distributions as the uncertainty range of the parameter.

Figure~\ref{fig:six} shows the Bayesian posteriors for the primary lens mass $M_1$ (upper panel) and 
$D_{\rm L}$ (lower panel).  In each panel, we present three distributions, in which the blue and red curves 
are the distributions contributed by disk and bulge lens populations, respectively, and the black curve is 
the sum of the two lens populations.  We find that the disk and bulge contributions are 23\% and 77\%, 
respectively.  The estimated masses of the primary and companion of the lens are
\begin{equation}
M_1 = 0.087^{+0.132}_{-0.047}~M_\odot,
\label{eq9}
\end{equation}
and 
\begin{equation}
M_2 = 0.19^{+0.28}_{-0.10}~M_{\rm J},
\label{eq10}
\end{equation}
respectively.  The planet mass is $\sim 0.63~M_{\rm S}$ in units of Saturn's mass, and thus the planet 
is a sub-Saturn planet.  We note that the mass of the primary approximately corresponds to the 
star/BD boundary.  Considering that the uncertainty of the estimated mass is substantial, the exact nature 
of the planet host is unclear.  The estimated distance to the lens is 
\begin{equation}
D_{\rm L} = 7.3^{+1.1}_{-1.1}~{\rm kpc}.
\label{eq11}
\end{equation}
The projected planet-host separation is
\begin{equation}
a_\perp = sD_{\rm L}\thetae = 0.62^{+0.09}_{-0.09}~{\rm au}.
\label{eq12}
\end{equation}
The planetary separation is much bigger than the snow line of $a_{\rm sl}\sim 2.7~{\rm au}(M_1/M_\odot)\sim  
0.23~{\rm au}$.  The estimated physical lens parameters of $M_1$, $M_2$, $D_{\rm L}$, and $a_\perp$ are 
listed in Table~\ref{table:four}.

% Table 4 ------------------------------------------------
\begin{table}[t]
\centering
\caption{Physical lens parameters\label{table:four}}
\begin{tabular*}{\columnwidth}{@{\extracolsep{\fill}}lc}
\hline\hline
\multicolumn{1}{c}{Parameter}     &
\multicolumn{1}{c}{Value}         \\  
\hline
$M_1$ ($M_\odot$)     &  $0.087^{+0.132}_{-0.047}$     \\
$M_2$ ($M_{\rm J}$)   &  $0.19^{+0.28}_{-0.10}$        \\
$D_{\rm L}$ (kpc)     &  $7.3^{+1.1}_{-1.1}$           \\
$a_\perp$ (au)        &  $0.62^{+0.09}_{-0.09}$        \\
\hline
\end{tabular*}
%\tablefoot{ ${\rm HJD}^\prime = {\rm HJD}- 2450000$.  }
\end{table}
% --------------------------------------------------------

We note that the estimates of the planet and host masses can vary depending on the assumption about the
planet hosting probability.  In our analysis, we assume that the planet frequency is independent of the 
host mass or the planet/host mass ratio.  \citet{Laughlin2004} discussed the possibility of the planet 
frequency dependency on the host mass by pointing out that, within the core accretion paradigm, giant 
planets would be difficult to be formed around low-mass M dwarfs, while such planets would be common 
around solar-type stars.  \citet{Vandorou2020} addressed this issue from a microlensing perspective by 
showing that the host mass of the planet MOA-2013-BLG-220Lb \citep{Yee2014}, with a planet/host mass 
ratio of $q\sim 3\times10^{-3}$, determined by analyzing the constraint from high-resolution AO 
observations was substantially more massive than the value estimated from the Bayesian analysis 
conducted under the assumption that stars of all masses were equally likely to host planets of a 
given mass ratio.  From this result, they suggested that planets with planet/host mass ratios 
$q\sim 2$--$3\times 10^{-3}$ might be more likely to be hosted by higher mass hosts.  Under this 
prediction, then, the mass of KMT-2018-BLG-0748L would be more massive than the mass presented in 
Equation~(\ref{eq9}).

\section{Discussion}\label{sec:six}

The planetary system KMT-2018-BLG-0748L illustrates that microlensing provides an important tool to 
detect planets orbiting very low-mass stellar and sub-stellar hosts.  To demonstrate the importance of 
microlensing in detecting such planetary systems, in Figure~\ref{fig:seven}, we present the distribution 
of planets in the plane of the host, $M_{\rm h}$, and planet, $M_{\rm p}$, masses for planetary systems 
with known masses.  There exist nine planetary systems, including KMT-2018-BLG-0748L, with host masses 
at around or below the star/BD mass limit of $\sim 0.08~M_\odot$.  These planetary systems include 
MOA-2007-BLG-192L \citep{Bennett2008, Gould2010, Kubas2012}, OGLE-2012-BLG-0358L \citep{Han2013}, 
MOA-2013-BLG-605L \citep{Sumi2016}, OGLE-2015-BLG-1771L \citep{Zhang2019}, OGLE-2016-BLG-1195L 
\citep{Bond2017, Shvartzvald2017}, KMT-2016-BLG-1107 \citep{Hwang2019}, OGLE-2017-BLG-1522L \citep{Jung2018}, 
and TRAPPIST-1 \citep{Gillon2016, Gillon2017}.\footnote{Here we exclude systems with $q\gtrsim 0.1$, because 
the large mass ratios suggest that they are more likely to form via the mechanism similar to that of binary 
stars rather than that of planetary systems \citep{Lodato2005}.} We mark these planetary systems by colored 
points: a red dot for KMT-2018-BLG-0748L, blue triangle dots for the other microlensing planetary systems, 
and green square dots for the seven planets in TRAPPIST-1 system. Among these systems, eight were detected 
using the microlensing method except for TRAPPIST-1, which was discovered using the transit method.

We note that the exact nature of the planet host can be identified from future high-resolution follow-up 
observations.  Considering that the mass of the planet host lies at around the star/BD boundary and the 
uncertainty of the estimated mass from the Bayesian analysis is large, it is not clear whether the host 
is a low-mass star or a BD.  Revealing the nature of the host is possible if the lens is resolved from 
the source by conducting high-resolution follow-up observations.  KMT-2018-BLG-0748 is an appropriate 
target for such observations because the relative proper motion, $\mu\sim 9.2~{\rm mas}~{\rm yr}^{-1}$, 
is substantially higher than typical lensing events.  \citet{Bennett2007} suggested the possibility of 
detecting the lens and source with high-resolution space-based follow-up observations, and this was 
realized by the detection of the lens star for the planetary microlensing event OGLE-2005-BLG-169 from 
the {\it Hubble Space Telescope} observations conducted when the lens-source separation was $\sim 49$~mas 
\citep{Bennett2015}.  The lens of the same event was also detected from the ground-based AO observations 
using the Keck telescope conducted by \citet{Batista2015}.  
Observations by the {\it James Webb Space Telescope}
({\it JWST}) would be able to probe much lower masses than observations by existing AO systems,
because the sensitivity of ground-based observations is limited by the bright sky background in the
infrared.
By implementing a criterion of $\sim 50$~mas 
separation for the lens and source resolution, the lens of KMT-2018-BLG-0748 could be resolved 
$\sim 5.4$~years after the discovery of the event, i.e, in early 2024.

However, in the case of KMT-2018-BLG-0748, one should adopt a significantly more conservative approach 
to future AO observations so that a non-detection would be unambiguously interpreted as due to the 
host being a sub-stellar object. That is, first, one should conservatively estimate the proper motion 
as $\mu > 7~{\rm mas}~{\rm yr}^{-1}$. And second, one should make allowances for the possibility of a 
large flux ratio between the lens and source due to the ultracool nature of the first, and so require 
$\Delta\theta > 70$~mas. Thus, to confidently exclude stellar-mass lenses in the event of a non-detection 
requires waiting $\Delta t = \Delta \theta/\mu > 10$~years, i.e., until 2028.

\section{Conclusion}\label{sec:seven}

The lensing event KMT-2018-BLG-0748 was analyzed and the results from the analysis were presented.
For the event, the central part of the light curve appeared to be asymmetric with respect to the 1L1S 
model due to the negative deviations in the rising side and the positive deviations in the falling side.  
We tested various models and found that that the deviations were explained by a binary-lens model with 
a mass ratio between the components of $q\sim 2\times 10^{-3}$.  The small angular Einstein radius, 
$\theta_{\rm E}\sim 0.11$~mas, indicated that the mass of the planet host was very low.  
From the Bayesian analysis conducted under the assumption of no dependency of the planet frequency on 
the host mass, it was estimated that the planet had a mass $M_{\rm p}=0.18^{+0.29}_{-0.10}~M_{\rm J}$ 
and the mass of the host, $M_{\rm h}=0.087^{+0.138}_{-0.047}~M_\odot$, was at around the star/BD boundary,  
but the host mass would vary depending on the assumption about the planet host probability. 
Resolving the lens and 
source would be possible in 2028, provided that the lens is luminous. In that case, the lens mass will 
be determined from these observations. If the measurement fails to detect the lens, this will imply that 
the lens is non-luminous and therefore, almost certainly, sub-stellar. In either case, these observations 
would resolve the nature of the planet host.

\begin{acknowledgements}
Work by CH was supported by the grants  of National Research Foundation of Korea 
(2017R1A4A1015178 and 2019R1A2C2085965).
% Gould  
Work by AG was supported by JPL grant 1500811.
% KMTNet
This research has made use of the KMTNet system operated by the Korea
Astronomy and Space Science Institute (KASI) and the data were obtained at
three host sites of CTIO in Chile, SAAO in South Africa, and SSO in
Australia.
\end{acknowledgements}


\end{document}